\documentclass[a4paper,twoside]{article}

\usepackage{graphicx}
\usepackage{epsfig}
\usepackage{subcaption}
\usepackage{calc}
\usepackage{amssymb}
\usepackage{amstext}
\usepackage{amsmath}
\usepackage{amsthm}
\usepackage{multicol}
\usepackage{apalike}
\usepackage{algorithm2e}
\usepackage[bottom]{footmisc}
\usepackage{graphicx}
\usepackage{svg}
\usepackage[hidelinks]{hyperref}
\usepackage{url} 
\usepackage{xurl} 
\usepackage{booktabs}
\usepackage{SCITEPRESS}     

\begin{document}

\title{4,500 Seconds: Small Data Training Approaches for Deep UAV Audio Classification}

\author{
    \authorname{Andrew P. Berg\sup{1}\orcidAuthor{0009-0000-1148-7174}, Dr. Qian Zhang\sup{2}\orcidAuthor{0000-0003-3166-4291}, Dr. Mia Y. Wang\sup{1}\orcidAuthor{0000-0003-2954-0855}}
    \affiliation{\sup{1}Department of Computer Science, College of Charleston, Charleston, SC, USA}
    \affiliation{\sup{2}Department of Engineering, College of Charleston, Charleston, SC, USA}
    \email{berga2@g.cofc.edu, zhangq@cofc.edu, wangy5@cofc.edu}
}


\abstract{ Unmanned aerial vehicle (UAV) usage is expected to surge in the coming decade, raising the need for heightened security measures to prevent airspace violations and security threats. This study investigates deep learning approaches to UAV classification focusing on the key issue of data scarcity. To investigate this we opted to train the models using a total of 4,500 seconds of audio samples, evenly distributed across a 9-class dataset. We leveraged parameter efficient fine-tuning (PEFT) and data augmentations to mitigate the data scarcity. This paper implements and compares the use of convolutional neural networks (CNNs) and attention-based transformers. Our results show that, CNNs outperform transformers by 1-2\% accuracy, while still being more computationally efficient. These early findings, however, point to potential in using transformers models; suggesting that with more data and further optimizations they could outperform CNNs. Future works aims to upscale the dataset to better understand the trade-offs between these approaches.}

\onecolumn \maketitle \normalsize \setcounter{footnote}{0} \vfill

\footnote{Code Repository \url{https://github.com/AndrewPBerg/UAV_Classification}}
\footnote{Experiment Tracking \url{https://wandb.ai/andberg9-self/projects}}
\footnote{Augmentation Notebook \url{https://colab.research.google.com/drive/1bl4RTQd7ENnMYEc4thwBwtocF-q1NYp2?usp=sharing}}

\section{\uppercase{Introduction}}
\label{sec:introduction}

Driven by affordability and rising popularity the use of commercial, military, and civil unmanned aerial vehicles (UAVs) is projected to increase in the next decade \cite{drone-stats}.  With this in mind, there is a need for a robust and highly accurate classification to prevent malicious and consequential security threats.

Common approaches to UAV classification include radio, visual, and acoustic approaches. As shown in \cite{modal24}: three common approaches are visual, radio frequency, and acoustic based. For each method, the constraining factor is the quantity and quality of the dataset. The inherent issue is that it is difficult to collect large amounts of UAV samples; for the simple fact that new military and consumer technologies advance yearly. Therefore, any solution to this problem needs to be able to cope with the scarcity of data.

This paper studies the difference in performance in audio classification models. We compare a custom convolutional neural network (CNN) against a pre-trained attention-based transformer. An important consideration is that the transformer model requires significantly more data to train from scratch.

This paper compares the transformer's results with a custom CNN model, using a scaled 9-class audio UAV dataset. To best compare the results, we opted for a fully attention-based transformer vs. a custom CNN. Because of the aforementioned data scarcity, training from scratch is not feasible on the larger transformer model, instead we opted to use the pre-trained Audio Spectrogram Transformer (AST) for this study. 

This paper builds on top of the work presented in \cite{Wang23}, in which a 22-class UAV audio dataset is presented. This dataset includes 100 samples of 5 second length each (11,000 seconds). For our purposes, we shift from 22 to 9 classes (4,500 seconds). Importantly, we chose to scale the dataset down to simulate the extreme data scarcity problem and to force any solution we test to overcome extreme data scarcity. Solutions use techniques like data augmentation and parameter efficient fine-tuning (PEFT), comparing the accuracies of the CNN and transformer models.

In later works we aim to answer at what quantity of data and with what techniques, will transformers be more viable than CNNs; and further which kind of training systems enable this with a small data problem?

We start by explaining our background knowledge, then move to show our methodology for tuning each model, and the techniques applied. Finally we explain the results of the tracked hyper parameter tuning and k-fold validation.

\section{\uppercase{Literature Review}}

Much work has been done in machine and deep learning on audio classification and detection problems \cite{audio-survey}.  CNN are proven to be the most effective model, and translate well to other domain specific audio sets. As shown in \cite{Wang_features}, a CNN approach works well on UAV audio classification. However, a key issue with any approach to UAV audio classification is the lack of data. A 22 class UAV audio dataset, as proposed in \cite{Wang23}, helps to alleviate this data gap, and studies the results of various machine learning (ML) and deep learning (DL) approaches. For this study we decided to scale down that dataset to 9 classes to better understand the effects of specific small data approaches shown later.

This paper also bases itself on acoustic data augmentation studies. \cite{sound_surveillance} and \cite{many_augs} prove the effectiveness of augmentations on model performance and outline the most relevant augmentations for UAVs.

Further, this paper approaches the modeling for small scale UAV classification using both CNN and attention-based transformer models \cite{attention-need}; comparing the best results from each.

A proof of concept approach for small datasets in the medical domain, using CNNs \cite{small-CNN}, shows precedent and demand for small dataset models; further motivating the small data approach.

 We propose that to make transformers work on such a small dataset, one must use a pre-trained transformer model. In this paper we use a pre-trained audio transformer and fine-tune to our custom datset. Fine-tuning models with a variety of PEFT techniques is a proven approach across application types; as shown in \cite{adapters-PLM} and \cite{PEFT-analysis}. This approach later proves that it is possible to make a pre-trained transformer competitive with the CNN's results.

\section{\uppercase{Methodology}}

\subsection{Tools}
\label{sec:tools}
The following lists all of the technologies important to the technical implementation of this study.
PyTorch, as introduced \cite{pytorch-paper}, is an essential deep learning framework for working with tensors, CUDA, and neural networks. This paper used it to create and modify the models, tensor datasets and iterable data loaders used throughout the experimentation. We used version 2.4.1 \cite{pytorch_docs}.

Torchaudio, is a domain-specific library based on PyTorch, for signal and audio processing with torch tensors. This paper used it to implement the data loading and parts of the feature extraction. We used version 2.4.1 \cite{torchaudio_docs}.

Weights \& Biases is a cloud-based machine/deep learning platform for robust experiment tracking and sharing. We used the package version 0.18.1 \cite{wandb_docs}.

Docker is a robust server tool used for containerized virtual environments for server-side deployment of code. We used it to talk with our local training server and maintain a consistent training environment. We used docker version 27.3.1 \cite{docker_docs}

Transformers is an open-source library to connect with Hugging Face's publicly hosted models, and integrate them into our code. As cited, it expands the amount of models available for all deep learning tasks \cite{transformers-paper}. We used version 4.44.2 \cite{transformers_docs}.

Parameter Efficient Fine-Tuning (PEFT) library is an open-source project from Hugging Face for abstracting the use of adapters. This was used to implement the majority of the adapters in the transformers experiments. We used version 0.12.0 \cite{peft_docs}.

Audiomentations is a open source library for data augmentation. This paper used the pre-built augmentations to process audio data and built custom augmentations. We used version 0.35.0 \cite{audiomentations}.

Other important python packages used in our code are: 
TorchMetrics\cite{torchmetrics_docs}, SciKit-learn\cite{sklearn_docs}, Librosa\cite{librosa_docs}, Matplotlib\cite{matplotlib_docs}, Telebot \cite{telebot_docs}, and NumPy\cite{numpy_docs}.

\subsection{Environment}

All training runs were performed on a local server with a AMD Theadripper 4950x, 128Gbs of RAM, and two NVIDIA GeForce RTX 4090s; both equipped with 24Gbs of VRAM. This was run on an Ubuntu Virtual Environment (version = 20.04). The server's local CUDA version was 12.2 and the virtual environment's was 12.1. All of the code is run in Python version.

\subsection{Custom UAV Dataset}
The dataset used in this paper is based on a 22 class custom dataset as presented in \cite{Wang23}. As of the time of writing, the dataset contains 31 data classes, each containing 100 instances of 5 second audio samples (15,500 seconds). For this paper's scope and scale we decided to use 9 of the total 31. That is 4,500 seconds of original audio data. Importantly, we decided not to study the full 31 class dataset in this paper to see what is possible on the subset and moving forward how our approach changes. The names and manufacturers of the UAVs are marked in Table \ref{tab:audio_classes}

\begin{table}[h]
\caption{UAVs included in the Dataset.}\label{tab:audio_classes}
\centering
\begin{tabular}{|c|c|c|}
  \hline
  Index & Manufacturer & Model\\
  \hline
  0 & Autel & Evo 2 \\
  1 & Self-build & David Tricopter\\
  2 & DJI & Avata\\
  3 & DJI & FPV\\
  4 & DJI & Matrice 200 \\
  5 & DJI & Mavic Air 2 \\
  6 & DJI & Mavic Mini 1  \\
  7 & DJI & Mavic Mini 2  \\
  8 & DJI & Mavic 2 Pro  \\
  
  \hline
\end{tabular}
\end{table}

\subsection{Data Exploration}
The data used in this paper is proprietary and private, meaning that this paper can only describe how the data is unique, rather than providing actual files. The UAV audio samples are continuous, with there being no natural silence in the sample. The samples are not clean, however; environmental noises like: wind, cars, people are included. 

\subsection{Feature Extraction}
\begin{figure}[ht]
    \centering
    \caption{Audio Analysis of DJI Tello Drone}
    \label{fig:audio-analysis}
    \includegraphics[width=0.4\textwidth]{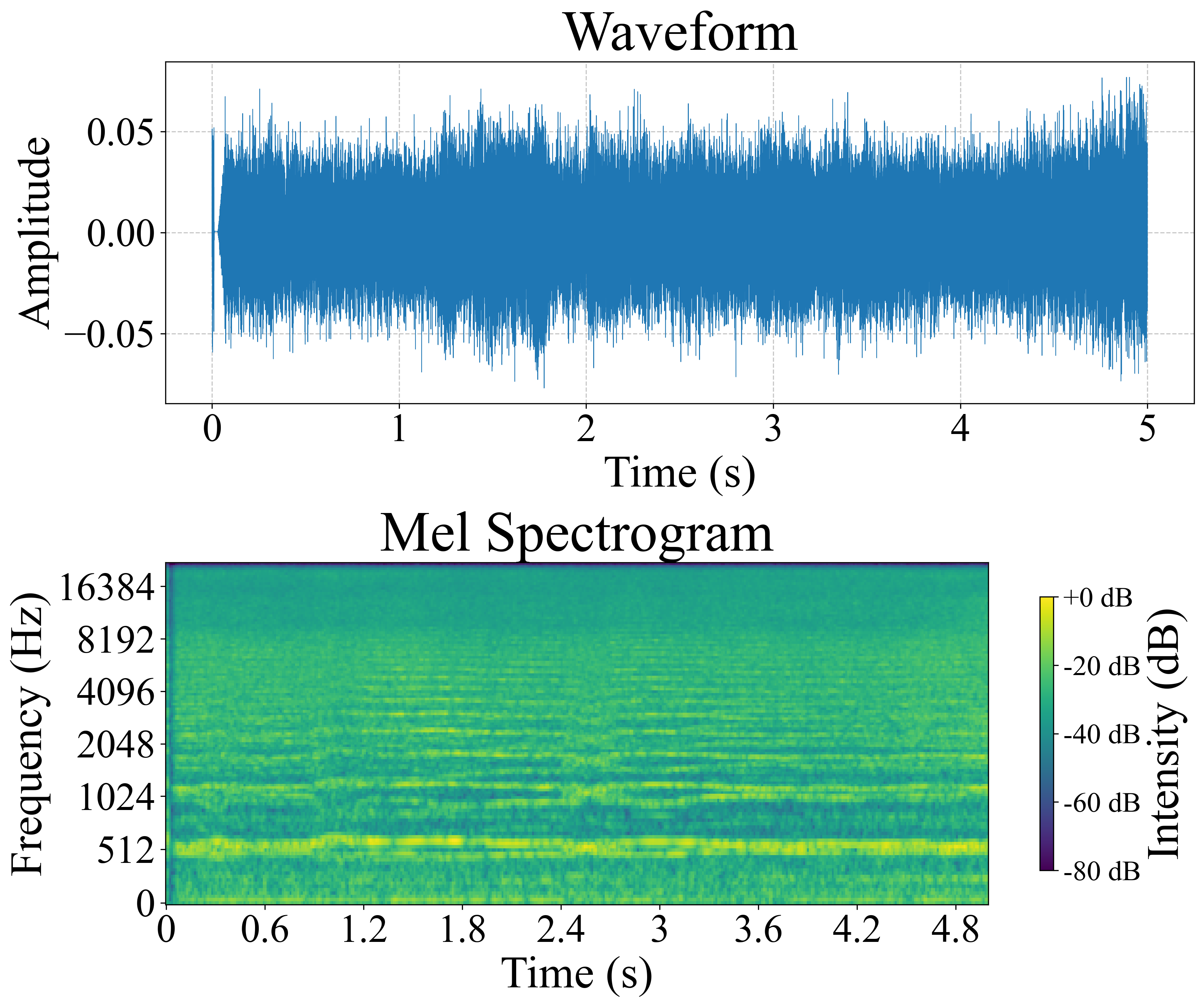}
\end{figure}

Feature extraction takes the complex and noisy waveform data and converts into a more compact, and more usable format. As illustrated in \ref{fig:audio-analysis}, both the custom CNN and pre-trained AST models use the mel-spectrogram feature extraction to capture the relative frequency content of the sample. Other popular options for audio feature extraction are MFCC, and log-mel-spectrograms. Interestingly, feature extracted audio samples function much like two-dimensional images.

\subsection{Data Standardization}
Some of the samples were recorded in mono and others dual-audio formats. Upon initialization all data is fixed so that every sample has one channel and a sample rate of 16000 Hz. If the audio is not exactly 5 seconds length, then the sample will either be padded or clipped to correctly fit into the model.

\subsection{Data Augmentation}

With such a limited number of samples, data augmentation is crucial for model performance and generalization. For our purposes we only augmented the training and validation runs. The correct combination of augmentations yield performance gains, specifically, inflating the training and train validation sets. Inflating the dataset with 3 augmentations per training and validation sample can add 2 times the original samples to the training and validation splits, further increasing generalization. Of course, more quality samples are always better than artificial ones, this approach is impractical for UAV datasets.

\subsection{Relevant Augmentations}

We found that the most relevant augmentations experiment for UAVs are: sin distortion, tanh distortion, time stretch, pitch shift, gaussian noise, and polarity inversion \cite{sound_surveillance}. We have compiled a Google Colab notebook for audio augmentation demonstration and visualization purposes \cite{colab_notebook}. Note, all data augmentations take place before feature extraction.

\subsection{Convolutional Neural Network}
\label{sec:CNN}

The convolutional neural network (CNN) implementation is based on \cite{Wang_features}. However, Where this paper's implementation diverges is in adding convolutional layers and favoring mel-spectrograms over the use of MFCC for feature extraction.

\subsection{CNN architecture}

Our custom CNN architecture has about 5 million parameters, using 3 convolutional layers and 2 fully connected layers after the convolutional passes. The models outputs a tensor with an output shape of 9. The model is visualized in figure \ref{fig:CNN-architecture}.

\begin{figure}[ht]
\caption{Custom CNN Architecture}
\label{fig:CNN-architecture}
\centering
\begin{small}
\begin{verbatim}
Total params: 5,006,825

TorchCNN(
  (conv1): Sequential(
    (0): Conv2d(1, 16)
    (1): ReLU()
    (2): MaxPool2d()
    (3): BatchNorm2d()
  )
  (conv2): Sequential(
    (0): Conv2d(16, 32)
    (1): ReLU()
    (2): MaxPool2d()
    (3): BatchNorm2d()
  )
  (conv3): Sequential(
    (0): Conv2d(32, 64)
    (1): ReLU()
    (2): MaxPool2d()
    (3): BatchNorm2d()
  )
  (fc1): Linear(19456, 256)
  (dropout): Dropout()
  (fc2): Linear(256, 9)
)
\end{verbatim}
\end{small}
\end{figure}

\subsection{Audio Spectrogram Transformer}
\label{sec:AST}
The Audio Spectrogram Transformer (AST): \cite{AST} breaks fully away from the use of convolutional layers in favor of a fully attention based architecture. Notably, AST is a fine-tuned version of the Vision transformer (ViT) \cite{VIT}, which was a landmark model which implemented a fully attention based model for image recognition.

AST is trained using cross-modality transfer learning. This is because the audio spectrogram and image data have similar formats. AST is trained on AudioSet \cite{AudioSet}, which is a collection of over 20 million seconds with 537 class labels. Classes include human sounds, animal sounds, natural sounds, music, sounds of objects, source ambiguous sounds, environment \& background noises.

Notably the AST author's express that the main disadvantage of a transformer is the need for more data to train on. Then AST was further fine-tuned and tested on ESC-50 \cite{ESC-50} and Speech Commands V2 Datasets \cite{SpeechCommandsV2}. AST achieved a precision of 95.6\% and 98.1\% using 5-fold cross-validation.

\subsection{AST architecture}\label{subsubsec:program_code}

This paper uses the MIT/ast-finetuned-audioset-10-10-0.4593 checkpoint hosted on Huggingface \cite{ast_model}. The model architecture is visualized in figure \ref{fig:AST-architecture}. 
First the model converts an input into embeddings. Next is the encoder layer which contains 12 transformer blocks, each with a self-attention, intermediate, and normalization layers. Finally after the transformer blocks, the model ends with a classifier head, which outputs a tensor with an output shape of 9.

\begin{figure}[ht]
\caption{AST architecture}
\label{fig:AST-architecture}
\centering
\begin{small}
\begin{verbatim}
Total params: 86,195,721

AST(
  (audio_spectrogram_transformer): ASTModel(
    (embeddings): ASTEmbeddings(
      (patch_embeddings):ASTPatchEmbeddings(
        (projection): Conv2d(1, 768)
      )
      (dropout): Dropout()
    )
    (encoder): ASTEncoder(
      (layer): ModuleList(
        (0-11): 12 x ASTLayer(
          (attention): ASTSdpaAttention(
            (attention):ASTSdpaSelfAttention(
              (query): Linear(768, 768)
              (key): Linear(768, 768)
              (value): Linear(768, 768 )
              (dropout): Dropout()
            )
            (output): ASTSelfOutput(
              (dense): Linear(768, 768)
              (dropout): Dropout()
            )
          )
          (intermediate): ASTIntermediate(
            (dense): Linear(768, 3072)
            (intermediate_act_fn): GELU()
          )
          (output): ASTOutput(
            (dense): Linear(3072, 768)
            (dropout): Dropout()
          )
          (LayerNorm): (768,))
          (LayerNorm): (768,))
        )
      )
    )
    (layernorm): LayerNorm
  )
  (classifier): ASTMLPHead(
    (LayerNorm): (768,))
    (dense): Linear(768, 9)
  )
)
\end{verbatim}
\end{small}
\end{figure}

\subsection{Feature Extraction}
AST takes audio mel-spectrograms as inputs, which is done using the Transformers library abstraction in the form of ASTFeatureExtractor.

\subsection{Model Fine-Tuning \& PEFT Techniques}

Because we are using a checkpoint of AST, we fine-tuned AST to our dataset. As opposed to the CNN approach. which we trained fully from random weights. Naive approaches to fine-tuning include full fine-tuning and classifier fine-tuning. Full fine-tuning is training all of a model's parameters and classifier trains only the final classifier. These are denoted as \textbf{Full} and \textbf{Classifier} in the results section.

In order to generalize the pre-trained weights we approached fine-tuning very carefully. Our leading technique to makeup for the lack of data quantity is to use parameter efficient fine-tuning. We implemented this in Hugging Face's PEFT library as referenced in the ~\ref{sec:tools}. This approach works by freezing all of the model's layers and only training added specialized adapter layers. 

\subsubsection{Low-rank Adaptation (LoRA)} 
LoRA allows a model to train some dense layers in a neural network indirectly by optimizing added rank decomposition matrices directly after the dense layers \cite{lora-paper}. LoRA layers are added evenly across the model. All while keeping the pre-trained weights frozen. Initially tested using Large Language Models (LLMs) like GPT-2 and RoBERTA; LoRA demonstrated similar performance to full fine-tuning while benefiting from significantly faster training time. This adds about 750,000 layers, where the LoRA rank = 8, and lora alpha = 16. These will be a part of the hyper-parameter tuning \ref{sec:tuning}.

\subsubsection{Adaptive Low-rank Adaptation (AdaLoRA)}
This is a new and improved version of LoRA proposed in \cite{adalora-paper} that adaptively budgets LoRA's layer "budget". Essentially, as the model is trained, only the most important layers are assigned with higher ranks and while the other non-important layers are de-ranked or even pruned. This adds more layers initially, but as described earlier will change during training. This allows us to set the rank and LoRA alpha much higher than computationally viable in a standard LoRA. For example, one could set the initial rank budget = 100 and the target rank to = 16 (and the LoRA alpha = 16) would start with 16,673,800 added layers, this is about equivalent if you were to set a standard LoRA with (r=100, a=16), however as described earlier, this will change in effect as the adaptive budget will reduce the rank over the course of the training.

\subsubsection{Orthogonal Fine-tuning (oft)}
Orthogonal literally means involving right-angles. Developed for text-to-image generation in mind, this fine-tuning approach uses orthogonal multiplication layers that sit at the ends of targeted blocks, then multiplying the output diagonally supposedly mimicking the angles in an image. \cite{oft-paper}. with a rank = 16 this approach adds about 9 million new layers. 

\subsubsection{Infused Adapter by Inhibiting and Amplifying Inner Activations (ia3)}
 Ia3 adds lightweight trainable layers after the model's attention and intermediate activation layers to efficiently rescale the model’s activation layers against a learned vector \cite{ia3-paper}. Initially proposed for LLMs, the founding ia3 paper demonstrates that the adapter had greater performance than LoRA adapters with even less added layers. While targeting the activation layers of the key, value, query, and dense attention layers ia3 only adds about 84,000 additional layers.

\subsubsection{Discrete Fourier Transformation Fine Tuning (Fourier FT)}
Fourier FT bases itself on the LoRA approach - adding low-rank structures - but instead learns a set of spectral coefficients using the Fourier transformation \cite{fourier-paper}. Effectively this compresses the amount of change resulting in less layers added to the model, which in turn leads to higher efficiency. With using a scaling = 100 and number of frequency coefficients = 3000 we can expect to add about 220,000 layers.

\section{\uppercase{Experiments \& Results}}

\subsection{Tuning}
\label{sec:tuning}

\begin{figure}[h!]
    \centering
    \includegraphics[width=0.47\textwidth, keepaspectratio]{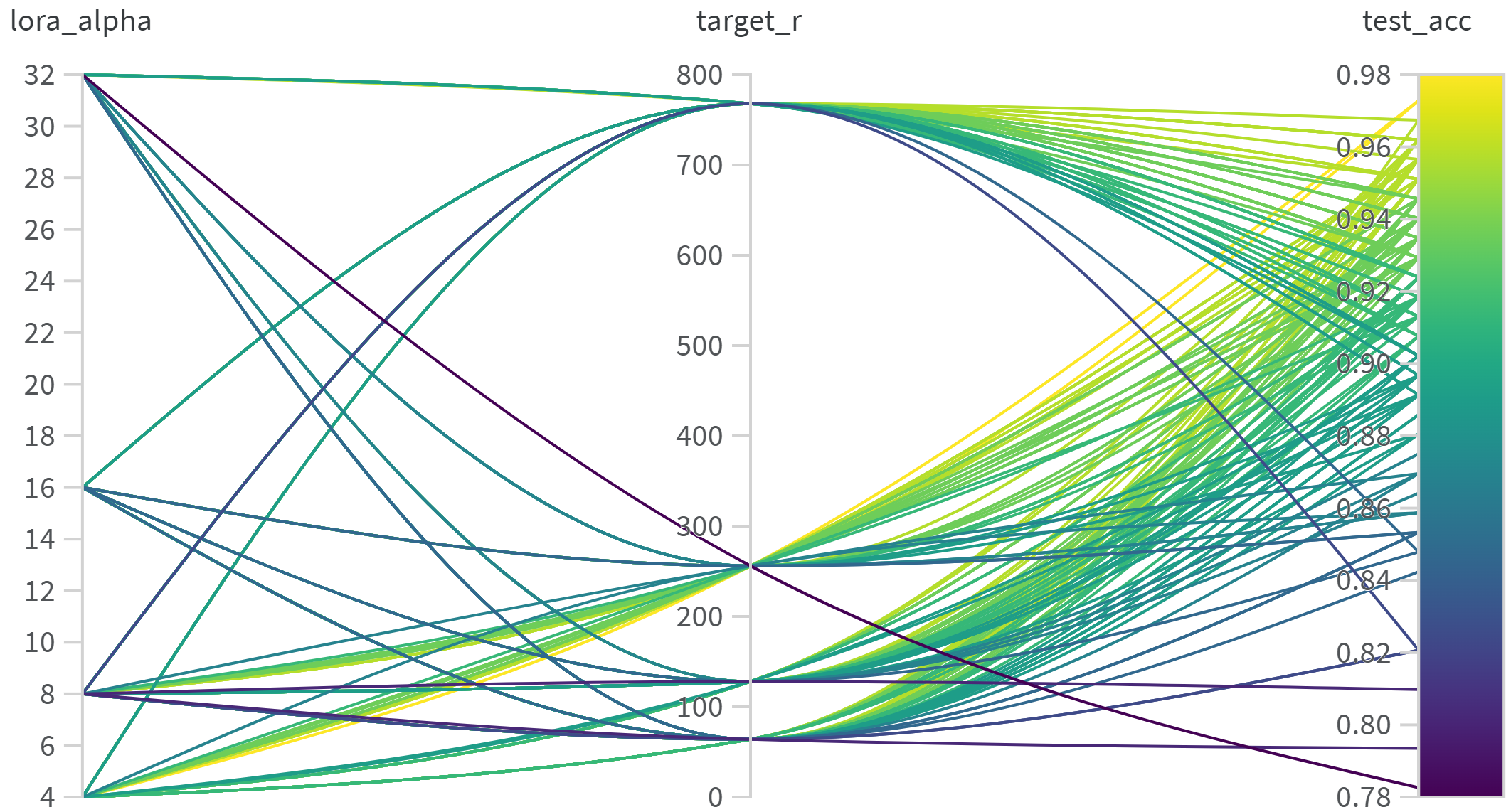}
    \caption{Example Weights \& Biases Hyper Parameter Tuning}
    \label{fig:h-tuning}
\end{figure}

With using a multitude of training and fine-tuning techniques, there was significant number of hyper-parameters to consider for tuning. For our experiment tracking and hyper-parameter tuning we used Weights \& Biases. The platform's sweep agents proved to be very useful in this task. As shown In figure \ref{fig:h-tuning}, a model (AST in the figure) can be trained given a range of selected hyper parameters and plotted to show the performance for each configuration. 

In this paper we tested a multitude of hyper parameters, such as: number of augmentations, type of augmentation, learning rate, type of adapters, etc. All of the results are tracked using Weights \& Biases and are publicly available projects as listed in \cite{wandb_projects}.

We used a 60-20-10-10 (train, test, train-val, test-val) split to evaluate the models in hyper-parameter tuning, and a 80-20 (train, test) split for 5-fold validation. We split the data, for hyper-parameter tuning, in this way to see the differences in augmentation vs. no augmentation data splits, so the train and the train-validation splits had augmentations applied, and the test and test-validation did not.

\subsection{Model Evaluation}
To evaluate the models we focused on the test accuracy and F1 score. Where F1 score is the harmonic mean of the evaluated precision and recall. These are standard classification metrics that allow for robust comparison of the training runs. Naturally, we also tracked the model's configuration, losses, precision, and recall; which can all be found on our Weights \& Biases projects page \cite{wandb_projects}.

\subsection{Results}

\subsection{AST PEFT}
To find the best possible PEFT technique for AST we trained the approaches given a range of configurations for each. We found that by far, the ia3 PEFT method had the most consistent validation accuracy, with discrete Fourier tuning and AdaLoRA close behind it. Please note that none of these sweeps included data augmentations, as to decrease the run time. The highest tested accuracies are listed in table \ref{tab:adapter-tuning}.
\begin{table}[h]
\small{

\caption{\textbf{AST PEFT Best results}}\label{tab:ast-results}
\centering
\begin{tabular}{l c c c}
\toprule
\textbf{PEFT Method} & \textbf{Accuracy(\%)} & \textbf{F1(\%)} & \textbf{Time$^\star $}\\
\midrule
Full        & 39.67          & 32.86 & 02:40            \\
Classifier  &\textbf{97.82} & 97.77 & \textbf{01:21}   \\
Ia3         &\textbf{97.83}  & 97.79 & \textbf{01:32}   \\
Fourier     & 94.02          & 93.90 & 02:44            \\
LoRA        & 53.26          & 48.68 & 03:39            \\
AdaLoRA     & 97.28          & 97.23 & 03:10            \\
Oft         & 95.10          & 95.14 & 06:30            \\
\bottomrule

\label{tab:adapter-tuning}
\end{tabular}
}
\end{table}
\footnote{$^\star$ Time is measured in (mm:ss).}

\subsection{Augmentation Experiments}
\label{sec:Augmentation-exp}
The purpose of this sweep of configurations was to find the optimal hyper-parameters for a given augmentation type. As shown in Table \ref{tab:aug-tuning} the three best augmentations tested are \textbf{sin-distortion, tanh-distortion, and time stretch}.

\begin{table}[h]

\caption{\textbf{AST Augmentations Best results}}\label{tab:aug-results}
\centering
\center
\begin{tabular}{l c c}
\toprule

\textbf{Augmentation} & \textbf{Accuracy(\%)} & \textbf{F1(\%)} \\

\midrule
control (no augs)   & 97.82          & 97.78 \\
sin-distortion      & \textbf{98.91} & 98.90 \\
tanh-distortion     & \textbf{98.91}  & 98.90 \\
pitch shift         & 97.28          & 97.22 \\
time stretch        & \textbf{98.91} & 98.87 \\
add noise           & 98.37          & 98.33 \\
polarity inversion  & 97.82          & 97.76 \\

\bottomrule
\label{tab:aug-tuning}
\end{tabular}
\end{table}

\subsection{Number of augmentations}
The aim of this experiment was to find the optimal number of augmentations per sample and the top 3 combination of augmentations as explained in Table \ref{tab:aug-results}, so either 2 or 3 augmentations would be selected for a run out of the selection pool of time stretch, sin distortion, and tanh distortion. 

The highest run recorded used sin distortion and time stretch augmentations, using the optimal respective augmentation configurations as determined in section \ref{sec:Augmentation-exp}, with 4 augmentations per sample. Training this run took 4 minutes and 19 seconds (not including the time taken to load \& augment the data) with a test accuracy of 99.45\% and a test F1 score of 99.44\%.

\subsection{CNN Tuning}
The purpose of these runs were to find the best learning rate and batch size for the custom pre-training of the CNN model outlined in section \ref{sec:CNN}. 

We tested batch size and learning rate without any augmentations.

The highest performing run took 34 seconds using a batch size of 8 (with 2 accumulation steps) and a learning rate of 0.001. This achieved a test accuracy and test F1 score of 98.88\%.  

When this configuration was later tested with varying number of augmentations per sample, it achieved exactly the same test F1 and accuracy with slightly higher runtime.

\subsection{5-fold Validation}
After tuning the models, we compared their average validation accuracies and F1 scores using a standard 5-fold validation approach. As shown in table \ref{tab:kfold-results}. On average the CNN model generally outperformed the AST model while having a significantly lower compute time.

\begin{table}[h]

\caption{\textbf{5-fold Validation Results (CNN \& AST)}}
\label{tab:kfold-results}
\centering
\begin{tabular}{l c c}
\toprule
  \textbf{Metrics} & \textbf{CNN} & \textbf{AST}\\
\midrule
  Best Accuracy, No Augs (\%) & 95.43          & 96.31 \\
  Best Accuracy, w/ Augs (\%) & \textbf{97.53} & 95.71\\
  Best F1, No Augs (\%)       & 95.48          & 96.21\\
  Best F1, w/ Augs (\%)       & 97.53          & 95.54\\
  Worst Time (mm:ss)          & 06:52          & 52:45\\
  Best Time (mm:ss)           & \textbf{01:30} & 06:19 \\
  Mean Time (mm:ss)           & 04:06          & 26:50  \\
\bottomrule  
\end{tabular}
\end{table}

\section{\uppercase{Conclusion}}
\label{sec:conclusion}

In this paper we have shown a process to fine-tuning and training models for extremely small audio datasets using CNN and transformer architectures.

Through careful testing the CNN approach is comparatively better than AST for UAV audio classification, given an extremely small 9-class dataset. However, given this fact, it is not fully understood at what amount of data or classes a transformer may be more effective than a standard CNN approach.

In future studies we would like to increase the amount of classes and audio data. Also, it would be interesting to use pre-trained CNN models like ResNet or MobileNet. Similarly, using other pre-trained transformer models would be good to gauge the effectiveness of augmentations and adapters, with a small chance of further increasing performance.

For practical applications of this research, a multi-modal embedded model would be a particularly useful for defensive compatibilities; with this, the model would have to be pruned and quantized for further efficiency.

\section*{\uppercase{Acknowledgments}}
This research is supported by the South Carolina Research Authority (SCRA), SCRA-Academic Collaboration Team Grant.

\bibliographystyle{apalike}
{\small
\bibliography{4500_seconds_paper}}

\end{document}